\documentclass[
 reprint,
showpacs,preprintnumbers,
 amsmath,amssymb,
 aps,
prb,
twocolumn,
]{revtex4}

\usepackage{graphicx}% Include figure files
\usepackage{dcolumn}% Align table columns on decimal point
\usepackage{bm}% bold math
\usepackage{natbib}
\begin{document}
\bibliographystyle{plainnat}

%\preprint{APS/123-QED}

\title{Muon diffusion in diluted paramagnetic systems}
\author{J. Rodriguez}
 \email{jose.rodriguez@psi.ch}
 \affiliation{Laboratory for Muon Spin Spectroscopy, Paul Scherrer Institut, 
	5232 Villigen, Switzerland}
\author{E. Pomjakushina}
 \affiliation{Laboratory for Developments and Methods, Paul Scherrer Institut, 
	5232 Villigen, Switzerland}
\author{A. Amato}
 \affiliation{Laboratory for Muon Spin Spectroscopy, Paul Scherrer Institut, 
	5232 Villigen, Switzerland}
\date{\today}

\begin{abstract}
Muon Spin Rotation/Relaxation ($\mu$SR) is a powerful probe 
to study diffusion of hydrogen-like particles. 
One of the most common model to analyze $\mu$SR signals from materials where
diffusion occurs, is by using functions derived from the stochastic 
theory of Kubo and Toyabe along with the strong collision model. Unfortunately
the current formalism can not be used to analyze data from systems where not
only muons diffuse, but also the magnetic moments of the
material are dynamic. In this article we introduce a new model which accounts
for both dynamical effects, and use this model to analyze $\mu$SR 
signals from $\mathrm{Dy_{0.4}Y_{1.6}Ti_2O_7}$. 
\end{abstract}

\maketitle

%\tableofcontents

\section{\label{intro}Introduction}

The study of muon diffusion is a topic of interest not only for science 
but also for technology, since the muon can be thought of as a light proton
\citep{storchak1998, alefeld1978}. Muon diffusion is measured
using the Muon Spin Rotation/Relaxation ($\mu$SR) technique. In this
technique polarized muons are implanted in a 
sample, and the positrons arising from their asymmetric decay are recorded
as a function of time and direction \citep{yaouanc2011}. 
With this information one may reconstruct the time dependent
polarization signal of the muon ensemble. The analysis of this signal
provides important information about the local magnetic environment and, 
if present, about the muon diffusion rate.

The most common model used to analyze $\mu$SR signals from systems
where muons diffuse is based in the probabilistic theory of 
R. Kubo and T. Toyabe \citep{kubo1967} together with the strong collision 
approximation \citep{kubo1967,hayano1979,uemura1985}. In its simplest
version this model  
assumes that, after the muons are implanted, they sense a local static
field which is random and has a Gaussian distribution. For paramagnetic
systems this local magnetic
field is typically originated by the nuclear magnetic 
moments, which are static within the $\mu$SR time window. The model
assumes also that the muons can hop between minima of potential 
with a hopping rate $\nu_h$, and therefore
the muons see instantaneous field changes with this same rate. 
This model, which is referred
as the Dynamical Gaussian Kubo-Toyabe (DGKT) model, is suited to analyze
signals from materials where the magnetic moments are 
densely packed. For magnetically diluted systems though, the internal 
field distribution is not Gaussian but Lorentzian
\citep{noakes1991,walstedt1974}.

No standard model exist to analyze signals if both, muon diffusion 
and fluctuations of the magnetic moments of the system, are present. 
As we will see in the
next section, in  magnetically dense systems a DGKT-like model
cannot disentangle muon hopping rates from fluctuations of the magnetic 
moments of the system. Nevertheless, this is not the case for 
magnetically diluted materials, and we report here a model suited to analyze
signals from this type of systems. 
Using numerical simulations of this model we have found that
magnetically diluted systems are specially well suited to study 
muon diffusion, since its onset produces a change
on the shape of the relaxing signal (even if the hopping rate is much smaller
than the fluctuation rate of the magnetic moments). After this, 
in Section \ref{pheno} we present a set of phenomenological formulas
derived from our simulated polarization functions, and which can be used 
to analyze experimental data. In Section \ref{experiment} we use these 
formulas to analyze data from $\mathrm{Dy_{0.4}Y_{1.6}Ti_2O_7}$; and
then in section \ref{comments}, we discuss the
results and comment on the limits of the model.

\section{\label{theory} A model for muon diffusion with magnetic
	moment fluctuations}

In the Gaussian Kubo-Toyabe (GKT) model \citep{uemura1999,yaouanc2011} 
the local field 
at each muon site is taken randomly from a Gaussian field 
distribution:

\begin{equation}
P_G(\vec{B},\Delta)=P^x_G(B_x,\Delta) \; P^y_G(B_y,\Delta) \;
	P^z_G(B_z,\Delta)
\label{e5}
\end{equation}

\noindent where:

\begin{equation}
P^i_G(B_i,\Delta)= \frac{\gamma}{\sqrt{2 \pi} \; \Delta} 
	\exp (-\gamma^2 B_i^2 / 2 \Delta^2) \
	\label{e1}
\end{equation}

In this last equation $\gamma$ is the muon gyromagnetic ratio 
(2$\pi \; \times$ 135.54MHz/T), $\Delta$ is the width 
of the field distribution, and
$B_i$ is one of the three components of the magnetic field. 
Since the local field at each muon site is static, then 
the polarization of the muon ensemble can be analytically
calculated and is given by:

\begin{equation}
A(t)=\frac{1}{3} + \frac{2}{3} ( 1 - \Delta^2 t^2) 
	\exp (-\Delta^2t^2/2)
	\label{e4}
\end{equation}

The GKT model describes systems of Heisenberg magnetic
moments which fill completely the sites of the lattice and
which are static.
If the field sensed by the muons changes
due to fluctuations of these magnetic moments, 
then the strong collision model can be used
to model the effect of these fluctuations on the muon polarization. 
In this strong collision
model \citep{uemura1999,yaouanc2011} the field change
is a Markovian processes with a characteristic
fluctuation rate $\nu_f$ 
({\it i.e.} after a time $1/\nu_f$ the
local field changes instantaneously to other 
taken randomly from the field distribution in Eq. 
\ref{e1}). The resulting dynamical model is the DGKT one
mentioned in the introduction. 
Now, if the local magnetic moments are static but the 
muons are diffusing at a rate $\nu_h$
(and assuming that the field at the 
new site where the muon hops is uncorrelated with the 
previous one), then the field
sensed by the muons will be effectively fluctuating with a rate
$\nu_h$, and the strong collision model can be also
applied to describe the muon diffusion process.
Then, in the DGKT model fluctuations of the 
magnetic moments are completely equivalent to a muon 
diffusion process; therefore, in magnetically dense systems 
it is not possible to distinguish one dynamic process from the other. 
This is not the case for magnetically diluted systems
as we will show in the following. 

Lets assume a static scenario 
again, but now most of the magnetic moments have 
been removed ({\it i.e.} substituted) from the lattice. In this diluted 
system the muons which
thermalize near to a magnetic moment can sense  higher
fields than those which land far from a magnetic moment. 
This means that many magnetically inequivalent muon sites 
are present in a diluted system. 
Mathematically, this can be model by introducing a distribution 
of $\Delta$. This distribution is given by \citep{uemura1981}:

\begin{equation}
P(\Delta,a)=\sqrt{ \frac{2}{\pi} }\; \frac{a}{\Delta^2} \;
	\exp ( -a^2/2\Delta^2 )
\label{e2}
\end{equation}

\noindent where $a$ is the characteristic width of the 
distribution. If the field distribution in Eq. \ref{e5} 
is averaged with 
the $\Delta$ distribution on Eq. \ref{e2}, the total magnetic
field distribution seen by the muon ensemble is 
the three dimensional generalization of the Lorentzian 
distribution as required for a diluted 
magnetic system \citep{noakes1991,walstedt1974}: 

\begin{equation}
P_L(\vec{B},a)=\frac{\gamma^3 a }
	{\pi^2 (a^2 + \gamma^2 B^2)^2}
\label{e3}
\end{equation}

\noindent where $B$ is the magnitude of the local field. 
The static polarization signal which arises from this
field distribution is:

\begin{equation}
A(t)=\frac{1}{3}+\frac{2}{3}(1-at)\exp (-at) 
\label{e10}
\end{equation}

\begin{figure}[tbp]
	\centering
	\includegraphics[width=\columnwidth,angle=0]{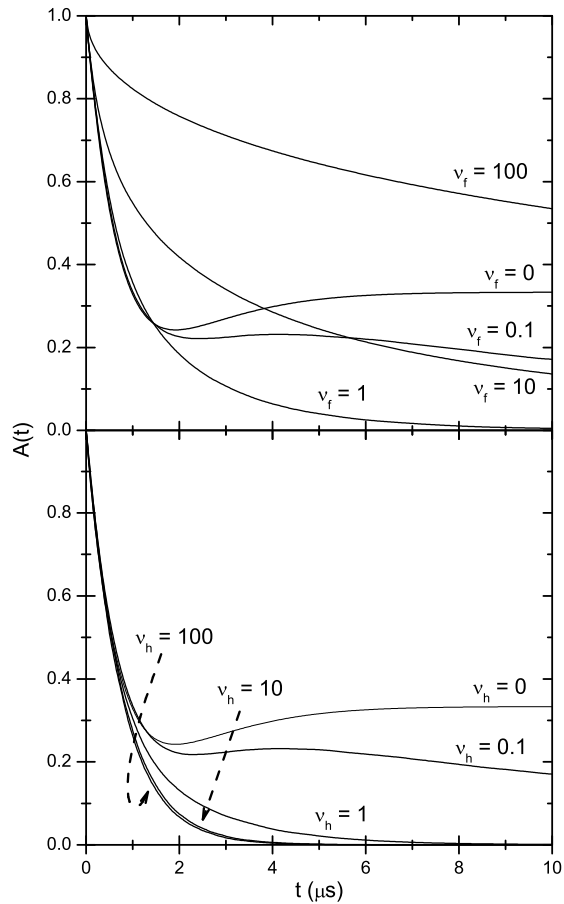}
\caption[] {Simulated signals for the FLKT model with $a$=1$\mu$s$^{-1}$
	and at different fluctuation rates
	of the local magnetic field (in $\mu$s$^{-1}$, upper panel);
	and for the HLKT model at different muon hoping rates
	(lower panel). The FLKT model shows the narrowing effect for 
	10$\mu$s$^{-1}$ and 100$\mu$s$^{-1}$, while the HLKT does not. 
	\label{f1}}
\end{figure}

If the 
field at the muon site changes due to fluctuations of the
magnetic moments of the material, then the 
dynamical behavior is introduced with the strong collision model 
\citep{uemura1999,yaouanc2011}. 
Note though, that muons which land in an specific region of 
the material will sense fluctuating fields with a distribution
characterized by a local $\Delta$; and
muons which land in other regions
will sense fluctuating local fields with a different local
$\Delta$. Mathematically
this amounts to making the time average (applying the strong 
collision model) before the 
spatial average (average over $\Delta$ with Eq. \ref{e2})
\citep{uemura1999,uemura1985}. $\mu$SR signals with this
Fluctuating Lorentzian Kubo-Toyabe (FLKT) model
are shown in the upper panel in Fig. \ref{f1} for different
fluctuation rates ($\nu_f$). We want
to notice two things. One is the strong depolarizing effect of 
dynamic fields (the polarization of the muon ensemble goes
to zero at long times, while it is equal to 1/3 in the static case); 
and second, the decrease in the relaxation
rate of the polarization signal at high fluctuation rates. 
This last effect is known
as the narrowing effect, and it happens because
the local field is fluctuating so fast that the muons do
not have time to precess much in the instantaneous local
field. An important characteristic of the signal in the 
narrowing limit is its square-root exponential 
trend\citep{uemura1999,yaouanc2011}:

\begin{equation}
A(t)=\exp (-4 a^2 t / \nu_f)^{1/2}
\label{e7}
\end{equation} 

Now, we will assume that the magnetic moments are static
but muons can hop from site to site. To account for this
dynamical effect
we have used the strong collision model again , but now 
we have performed the time average after the spatial average. 
This means that a single muon can probe the whole 
distribution in Eq. \ref{e3} as it hops through
the lattice (assuming again that the field from site to 
site is uncorrelated), and not
only one specific $\Delta$. Then, 
muons are able to sense a bigger range of fields 
as they diffuse faster and faster. The lower panel in
Fig. \ref{f1} shows
several muon polarization signals using this Hopping 
Lorentzian Kubo Toyabe (HLKT) model. Notice that
no narrowing effect is observed. Further more,
at high hopping rates the signal is exponential and
independent of the hopping rate \citep{fiory1981,uemura1981}:

\begin{equation}
A(t)=\exp (-4 a t/3)
\label{e6}
\end{equation}

The absence of the narrowing effect in the HLKT model
is a problem of the Lorentzian field 
distribution, which allows for the muon to sense 
unphysically large magnetic fields \citep{fiory1981}. To
appreciate this, we have 
calculated the probability that a muon senses a field
bigger than $\nu_h / \gamma$ when hoping into a new
site (at these fields the muons
will be able to make a big precession before hoping
to an other site and therefor reduce the polarization of 
the ensemble). At high hoping rates ($\nu_h \gg a$) this 
probability is $4a/\pi \nu_h$. This can be used to 
calculate the probability that within the signal time 
window (typically 10 $\mu$s) the muon sense a 
highly depolarizing field. This probability is equal 
to $4Ta/\pi$ ($T$ is the signal time window) and it is 
independent of $\nu_h$. This means that as $\nu_h$ is 
increased more and more the signal will tend to a 
$\nu_h$ independent relaxation rate. 
This unphysical behavior of the HLKT model could be 
avoided by introducing a maximum field cutoff
in the distribution on Eq. \ref{e3}.
This cutoff is system dependent and should
be calculated carefully. This type of refinement of 
the model is outside the scope of this document. Never
the less, the HLKT model is valid as long as 
$\nu_h \lesssim a$.

We have extended the Lorentzian Kubo-Toyabe 
formalism to account for systems where the magnetic 
moments fluctuate and also 
muons diffuse. In this model, which we will 
call the Extended Lorentzian Kubo-Toyabe (ELKT) model,
the dynamical
behavior is also introduced by the strong collision
model, but this time before and after the spatial 
average to account for the magnetic moment fluctuations 
and muon hoping respectively. We have studied
this model numerically using a Monte Carlo algorithm.
At time zero, this algorithm selects a $\Delta$ for the 
local field distribution with a probability given by
Eq. \ref{e2}. Then, the local field is allowed to 
to fluctuate with a fluctuation
rate $\nu_f$. The local field fluctuations are 
carried on until a muon-hop 
event happens, and this occurs with a frequency given 
by the muon
hopping rate $\nu_h$. At this point a new $\Delta$ is 
randomly chosen from the probability distribution on
Eq. \ref{e2} and, again, the local field is allowed
to fluctuate
with a field distribution characterized by the new 
field width. In between all these ``collisions" the 
muon polarization function is properly evolved, and it
is recorded into an array at a discrete set of times. 

\begin{figure}[tbp]
	\centering
	\includegraphics[width=\columnwidth,angle=0]{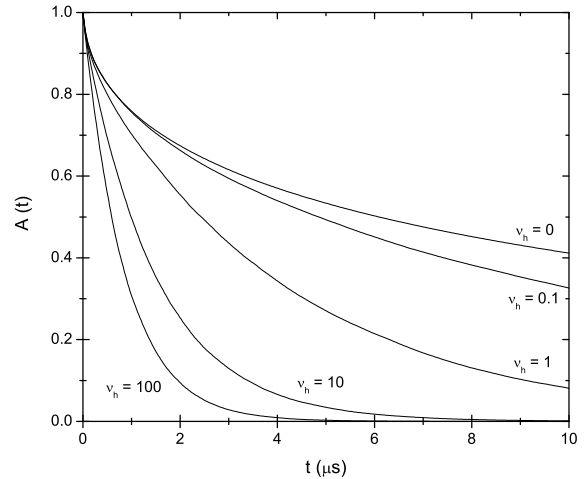}
\caption[] {Simulated signals for the ELKT model with $a$=1$\mu$s$^{-1}$
	and $\nu_f=50\mu$s and for different muon hopping rates (in 
	$\mu$s). The relaxation of the signal increases continuously
	with $\nu_h$ and the signal changes from a square root exponential
	to a simple exponential continuously. 
	\label{f2}}
\end{figure}

Since our interest is to study the effect of
muon diffusion in fast fluctuating paramagnetic materials 
(see Section
\ref{experiment}), we have performed all our simulations
in the narrowing limit ($\nu_f \gg a$). Fig. \ref{f2}
shows the effect of increasing the muon hopping rate
on the $\mu$SR signal from a diluted 
paramagnetic system. The signal is a root exponential 
in the absence of muon diffusion and equal to 
Eq. \ref{e7}. As the muon hopping rate is increased,
the signal shape changes continuously to an exponential 
at high diffusion rates. In the high hopping rate
limit the signal is given by Eq. \ref{e6}. 

\section {\label{pheno}Phenomenological description of the ELKT model}

In order to use the numerical solutions of the ELKT model
to analyze experimental data 
(see next section), we have fitted our simulated signals in the 
narrowing limit ($\nu_f \gg a$) to a stretched 
exponential function:

\begin{equation}
A(t)=\exp (-(\lambda t)^\beta)
\label{e11}
\end{equation}

Fig. \ref{f3} shows the dependence of $\beta$ and $\lambda$
on the muon hopping rate. Note that, as the muon hopping rate is 
increased from zero, the parameter $\beta$ grows from 0.5 to 1
as expected.
We have found that the dependence of $\beta$ and $\lambda$ on
the field fluctuating rate, muon hopping rate and $a$ is well 
described by (see Fig. \ref{f3}): 

\begin{figure}[tbp]
	\centering
	\includegraphics[width=\columnwidth,angle=0]{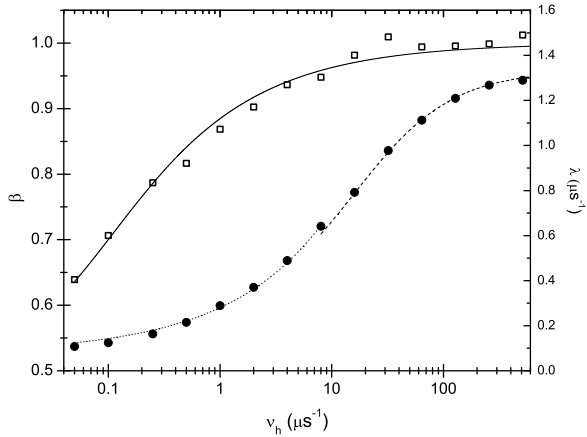}
\caption[] {Values of $\beta$ (squares, left y-axis) and $\lambda$ (circles,
	right y-axis) as a function of $\nu_h$. These 
	values were found from fitting the simulated data,
	with $a=1\mu$s$^{-1}$ and $\nu_f=50\mu$s$^{-1}$, to Eq. \ref{e11}.
	Lines are the plots of Eq. \ref{e12} (continuous line), \ref{e13}
	(dotted line) and \ref{e14} (dashed line).
	\label{f3}}
\end{figure}

\begin{equation}
\beta= 1 - \frac{1}{2} \left( \frac{3.9}{\displaystyle \frac{\nu_h}{a} \left(
	\frac{\nu_f}{a}+20.1 \right) + 3.9} \right)^{1/2}
\label{e12}
\end{equation}

\begin{equation}
\lambda_{low} = \frac{4a^2}{\nu_f} + a \left( \frac{\exp (-\nu_f/90a)}
	{5.4} + \frac{1}{10.6} \right) \sqrt{\frac{\nu_h}{a}}
\label{e13}
\end{equation}

\begin{equation}
\lambda_{high} = a \left( \frac{4}{3} - \frac{0.35}{\displaystyle 
	\frac{\nu_h}{\nu_f} + 0.32} \right)
\label{e14}
\end{equation}

\noindent where $\lambda_{low}$ and $\lambda_{high}$ refer to 
different ranges where each equation apply. $\lambda_{low}$
describes the $\nu_h$ dependence for $\nu_h < 8a$, and 
$\lambda_{high}$ for $\nu_h > 8a$.

We want to mention that, as a log(log($A$)) 
vs. log(t) plot shows, the numerical solutions of the ELKT
model are close to stretched exponentials 
but are not equal. Nevertheless we have used this function
because is commonly used in the analysis of $\mu$SR signals
and it will allows us to understand previous published 
works (see Section \ref{comments}). 
In the next section we 
use Eq. \ref{e12} and \ref{e13} to obtain 
quantitative measurements of 
the muon diffusion rate in $\mathrm{Dy_{0.4}Y_{1.6}Ti_2O_7}$.

\section{\label{experiment}Muon diffusion in $\mathrm{Dy_{0.4}Y_{1.6}Ti_2O_7}$ }

The low temperature behavior of the ionic insulator $\mathrm{Dy_2Ti_2O_7}$ 
have had much attention from researchers \citep{gingras2011}. Below 20K, 
the Dy$^{3+}$ magnetic
ions have an Ising character and they form a magnetically frustrated system
due to their effectively antiferromagnetic interactions and to their
arrangement in a lattice of corner shearing tetrahedra. 
Below $\sim$4K, the magnetic ions arrange such that two spins point
inside every tetrahedra and two out. This arrangement of spins can be 
mapped into that of hydrogen atoms in watter ice, and therefore
this ground state is often referred as spin ice. More recently, it was 
noticed that the spin ice state is a background
for topological excitations which behave as magnetic monopoles
\citep{castelnovo2008,jaubert2009}.  

We have used $\mu$SR to study the effect of magnetic dilution in this
system by substituting magnetic Dy$^{3+}$ ions with non magnetic 
Y$^{3+}$. A single crystal of $\mathrm{Dy_{0.4}Y_{1.6}Ti_2O_7}$ was
grown using an optical floating zone furnace. The 
single crystal was aligned using a Laue camera and cut such that the
muons are implanted with their spin parallel to the (110) direction. All 
the sample preparation was performed in the Laboratory for Developments
and Methods in the Paul Scherrer Institut (PSI, Switzerland). The 
$\mu$SR measurements were performed at the GPS instrument at PSI, and which
is located in the $\pi$M3 beam line. We
performed temperature scans between 2K and 250K 
in an applied field of 100G transverse to the muon polarization. Also,
we have performed complementary measurements with the field
applied along the muon spin direction and in zero field.

\begin{figure}[tbp]
	\centering
	\includegraphics[width=\columnwidth,angle=0]{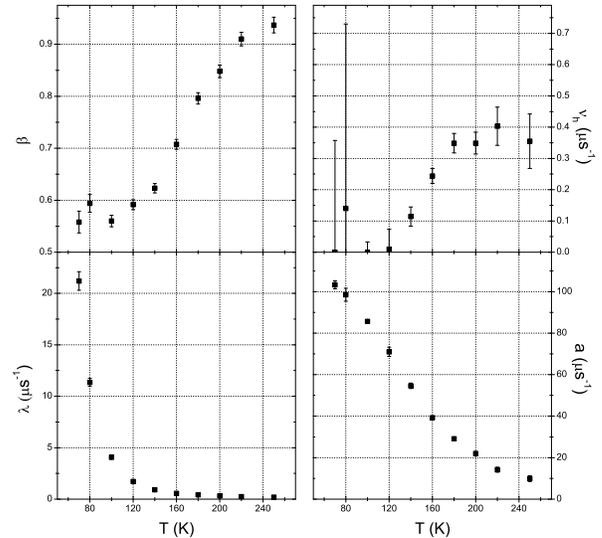}
\caption[] {Left panel: coefficient of the power exponential relaxing 
	envelope ($\beta$) and relaxation rate of the signal ($\lambda$), 
	from fitting the data of $\mathrm{Dy_{0.4}Y_{1.6}Ti_2O_7}$
	in zero external field with Eq. \ref{e11} times a cosine function.
	Right panel: muon hopping rate ($\nu_h$) and characteristic width 
	of the field distribution ($a$) as a function of temperature;
	obtained by fitting the data with $\lambda$ and
	$\beta$ given by Eq. \ref{e12} and \ref{e13} respectively.
	\label{f5}}
\end{figure}

Fig. \ref{f4} shows a $\mu$SR signal from $\mathrm{Dy_{0.4}Y_{1.6}Ti_2O_7}$
at 160K where the system is paramagnetic. Note in this picture the stretched 
exponential-like envelope of the signal. For this reason, we first
fitted the data
to a cosine function times a stretched exponential envelope. 
The left panel in Fig. \ref{f5} shows that
the parameter $\beta$ increased with temperature from a 
value of $\approx$0.59 below 100K to $\approx$0.94 at 250K, and
$\lambda$ decreased monotonically from 21$\mu$s$^{-1}$ to 
0.2$\mu$s$^{-1}$ in the same range. 
As we mentioned in the previous section, the ELKT model 
can explain the trend in $\beta$ in terms of an 
onset of muon diffusion upon warming. Nevertheless, and according to 
Fig. \ref{f3}, an onset of muon diffusion would produce an 
increase of $\lambda$ with temperature. This is in contrast with what
we observe. As we will see later, the decrease of 
$\lambda$ with temperature is also captured by the ELKT model,
and it is due to the fact that the thermally activated
fluctuations of the Dy$^3+$ magnetic moments are much bigger than
the muon hopping rate (see Eq. \ref{e13}).
Then, we analyzed our data by fitting the
relaxing envelop to a stretched exponential where the relaxation
rate is given by Eq. \ref{e13} and the coefficient 
by a modified version of Eq. \ref{e12}. This modification
consist on substituting  
the 1/2 factor on the right hand side of Eq. \ref{e12} by 0.41, 
such that $\beta$ for T$<$120K is equal to the 
experimentally observed value of 0.59 . This apparently {\it ad hoc} 
modification is justified by the fact that, even if  
the dilution is significant in our system, it is not in the
very high dilution regime which the ELKT model assumes (20\% of 
the DY/Y sites are occupied by magnetic ions in 
$\mathrm{Dy_{0.4}Y_{1.6}Ti_2O_7}$). As shown
by Noakes\citep{noakes1991}, as a system is magnetically
diluted the relaxation function changes continuously from a
Gaussian Kubo-Toyabe to a Lorentzian Kubo-Toyabe. This
means that in the narrowing limit, the coefficient of
the stretched exponential will change from 1 to 0.5 
continuously as the system is magnetically diluted. Our
system is concentrated enough, such that this coefficient
is slightly above 0.5 

It is well known that in $\mathrm{Dy_2Ti_2O_7}$ the 
characteristic fluctuation rate of the magnetic
moments in the paramagnetic state
is thermally activated \citep{snyder2001,ehlers2003}, and 
therefor the fluctuation rate of the local field at the
muon site will be given by\citep{uemura1985}:

\begin{figure}[tbp]
	\centering
	\includegraphics[width=\columnwidth,angle=0]{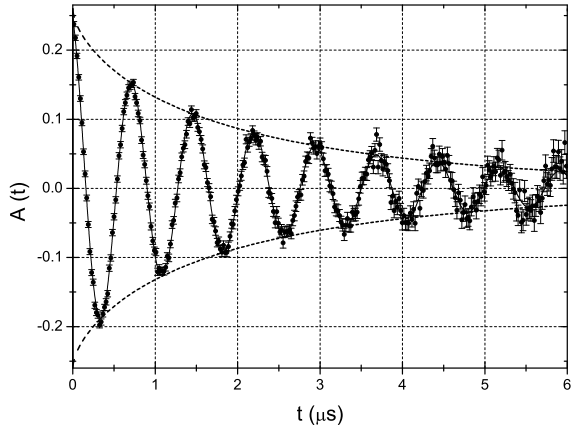}
\caption[] {$\mu$SR signal from $\mathrm{Dy_{0.4}Y_{1.6}Ti_2O_7}$ at 160K
	and in a transverse field of 100G. The continuous line is a fit
	to Eq. \ref{e11} multiplied by the experimental asymmetry 
	(amplitude) and a cosine term. The dashed lines are the fitted
	envelope of the signal ({\it i.e.} without the cosine term).
	\label{f4}}
\end{figure}

\begin{equation}
\nu_f=\nu_0\exp (-E/k_BT)
\label{e15}
\end{equation}

\noindent where $E$ is the energy gap between the ground state
doublet and the first excited crystal field level. The reported
values for this number ranges from 200K to 380K 
\citep{snyder2001,rosenkranz2000}. In the absence of spectroscopic
data, we used a value equal to 300K. In order
to determine $\nu_0$ we used a longitudinal field measurement
at 60K to determine the fluctuation rate of the local magnetic
field at this temperature. Assuming that the muons do not diffuse
at these temperature (this is indicated by a temperature independent 
$\beta$ of approximately 0.59), we have obtained a fluctuation 
rate of 990MHz at this temperature. Using this value in 
Eq. \ref{e15}, we have obtained $\nu_0$=150GHz,
which is close to that found by neutron spin echo measurements
(220GHz) \citep{ehlers2003}. Then, when we analyzed our 
data we set $\nu_f$ to the value given by Eq. \ref{e15}
and extracted $a$ and $\nu_h$ from the fit at each temperature. 

The fitted values for $\nu_h$ and $a$ are shown in the right panel
in Fig. \ref{f5}. Note the big error bars on $\nu_h$ for 
T$\leq$80K. This only reflects the very steep grow of $\beta$
at low values of $\nu_h$ (see Eq. \ref{e12}). Figure
\ref{f5} shows that muons start to diffuse between 
120K and 140K, and the diffusion rate grows until 
$\approx$200K and then stays constant. Below 80K, the 
parameter $a$ is constant and equal to 100$\mu$s$^{-1}$.
This is close to 110$\mu$s$^{-1}$, value that we found 
at 2K where the system is almost static within our time 
window. As the temperature grows the value of $a$ decreases
due to the different orientations that the magnetic moments can
have in the higher crystal field level. 

\section{\label{comments}Comments}

As shown in the previous section, the ELKT model 
is able to explain the increase of $\beta$ with temperature 
in $\mathrm{Dy_{0.4}Y_{1.6}Ti_2O_7}$
in terms of the onset of muon diffusion. This increase of 
$\beta$ with temperature has been
observed in an other ionic insulator
\citep{rodriguez2009,johnson2011}, but no explanation of this
effect was provided. Campbell {\it et al.} \citep{campbell1994}
observed an increase 
of $\beta$ with temperature in the canonical spin-glass {\it Ag}Mn at
high temperature \citep{spinglass}. The authors mentioned that this
effect could be produced by muon diffusion but no further analysis of
the data was performed. This study also pointed out that
the increase in $\beta$ was accompanied
by an increase, and subsequent saturation, of the relaxation rate of
the signal upon cooling. The authors interpreted this as a consequence of muon
retraping near the magnetic impurities. In our case the 
relaxation rate of the signal decreases monotonically with 
temperature up to 250K. 
Nevertheless, the ELKT model shows
a saturation of $\nu_h$ above 160K (see Fig. \ref{f5}). 
It is not possible to know with the data taken 
if this saturation is due to muon
retraping (the slightly different ionic radius
of Y and Dy create local distortions in the lattice \citep{snyder2004}
which might favor some muon sites over others). 
Nevertheless, we want to note that if a 
muon retraping process is present in $\mathrm{Dy_{0.4}Y_{1.6}Ti_2O_7}$,
then the ELKT model would provide only with an ``averaged" 
muon diffusion rate since this model posses only one diffusion
parameter (instead
of the three dynamical parameters needed to describe diffusion 
with a single retraping mechanism\citep{yaouanc2011}). Still, and even 
if muon retraping is present, the increase in $\beta$ from
0.59 to 0.94 remains a signature of the onset of muon diffusion
since its change is produced by the muon exploring different regions 
of the material with different magnetic moment densities 
({\it i.e.} different characteristic field widths $\Delta$).
This fact is confirmed by our 
low transverse field measurements in a sample of  
$\mathrm{Dy_{1.6}Y_{0.4}Ti_2O_7}$, where we found that the 
parameter $\beta$ is temperature independent 
and equal to 1 within error bars. This is expected by the 
theory since this sample is magnetically dense, and therefore
all muon sites have approximately same local density of
magnetic moments ({\it i.e.} the same field width $\Delta$).

We want to finish this section by reviewing a couple of limitations of 
the ELKT model. One
limitation is the spherical nature of the internal field 
distribution which the ELKT model assumes (see Eq. \ref{e5}). 
Care should be taken when
using Kubo-Toyabe models to fit single crystal data since the 
internal field distributions have not, in general, spherical
symmetry. Nevertheless 
the biggest discrepancies are expected to be observed
at low values of $\nu_f$ and not so much at
the high fluctuation rates explored in this work \citep{rodriguez2009}. 
An other limitation of
the ELKT is that it assumes that the magnetic fluctuations are 
characterized by a single fluctuation rate of the local magnetic
field ($\nu_f$). Then,
it should not be used to model magnetic systems with a richer dynamic
response such as spin-glasses relatively close to $\mathrm{T_g}$.

\section{\label{conslussions}Conclusions}

We have introduced the ELKT model which is suited 
to analyze $\mu$SR signals from diluted
magnetic systems where muon diffusion is present and where the magnetic 
moments are dynamic. Also, we have provided
a set of phenomenological formulas to extract muon diffusion 
rates from systems where the magnetic background fluctuates fast. We have used
these formulas to analyze data from $\mathrm{Dy_{0.4}Y_{1.6}Ti_2O_7}$, and
found that muons diffuse for T$>$140K. Also, we have found that the 
diffusion rate saturates for T$\ge$180K. We have shown that the 
increase with temperature of the parameter $\beta$ in a hot diluted paramagnet
is produced by the onset of muon diffusion, and this is further
confirmed by our observation that this 
parameter is temperature independent in a magnetically dense system.
We believe that the use of the ELKT model can allow to study muon 
diffusion by introducing a small amount of strong magnetic impurities 
in a system where muon diffusion can not be studied otherwise
({\it e.g.} in systems with small nuclear magnetic moments).
 
%NOTE: adjust this temperature. 

\bibliography{dytio}% Produces the bibliography via BibTeX.

\begin{thebibliography}{22}
\bibitem[{spi()}]{spinglass}
\bibinfo{note}{The increase in $\beta$ explained in this document should not be
  confused with that observed in spin-glasses as one warms up from
  $\mathrm{T_g}$. This last one is produced by a broadening of the dynamic
  response of the system at low temperature, while that presented in this
  document is due to muon diffusion.}
\end{thebibliography}

\end{document}